\documentclass[prl,twocolumn,showpacs,floatfix,preprintnumbers,amsmath,amssymb,superscriptaddress]{revtex4}

\usepackage{graphicx}
\usepackage{amsmath}
\usepackage{amssymb}
\usepackage{color}
\usepackage{dcolumn}
\usepackage{epsfig}
\usepackage{bm}
\usepackage{subfigure}
\usepackage[urlcolor=blue]{hyperref}
\hypersetup{backref, colorlinks=true, linkcolor=blue, citecolor=blue}

\begin{document}

\title{Superconductivity in $p$-Terphenyl}

\author{Ren-Shu Wang}
\affiliation{Center for High Pressure Science and Technology Advanced Research, Shanghai 201203, China}
\affiliation{School of Materials Science and Engineering, Hubei University, Wuhan 430062, China}

\author{Yun Gao}
\email{gaoyun@hubu.edu.cn}
\affiliation{School of Materials Science and Engineering, Hubei University, Wuhan 430062, China}

\author{Zhong-Bing Huang}
\email{huangzb@hubu.edu.cn}
\affiliation{Faculty of Physics and Electronic Technology, Hubei University, Wuhan 430062, China}

\author{Xiao-Jia Chen}
\email{xjchen@hpstar.ac.cn}
\affiliation{Center for High Pressure Science and Technology Advanced Research, Shanghai 201203, China}

\date{\today}

\begin{abstract}
Motivated by the exploration of bipolaronic superconductivity in conducting polymers, we examine such a possibility in the starting member of $p$-oligophenyls $-$ $p$-terphenyl with three phenyl rings, belonging to the family of conducting polymer polyparaphenylene. The formation of bipolarons is identified from Raman scattering measurements. Both the dc and ac magnetic susceptibility measurements reveal that $p$-terphenyl is a type-II superconductor with a critical temperature of 7.2 K upon doping potassium. The electron-lattice interaction, manifested by pronounced bipolaronic bands, is suggested to account for the observed superconductivity. Conducting polymers are thus demonstrated to have the potential for the development of new superconducting technologies and devices. 
\end{abstract}


\maketitle

Conducting polymers have been the focus of a great deal of research in the past nearly 40 years \cite{1}. These materials possess the excellent combination of easy processability, light weight, and durability of plastics with the high electrical conductivity for technological applications. Meanwhile, they also helped to spur new areas of fundamental research in condensed matter physics by demonstrating some exotic physical properties \cite{2}. Bipolarons and polarons, $i.e.$, self-localized quantum states associated with characteristic distortions of the polymer backbone, are thought to play the leading role in determining the charge injection, optical, and transport properties of conducting polymers \cite{3,4,5,6,7,8}. Experiments \cite{6,7,8} and theories \cite{3,4,5} have established that doping these polymers in the low and high levels result in the formation of paramagnetic polarons and diamagnetic bipolarons, respectively. It has been recognized that the bipolaron system is more stable than the polaron one when dopants are taken into account, which is manifested by a slow evolution of polarons into bipolarons \cite{7,8}. The appearance of bipolaronic bands offers a natural explanation for the vanishing small Pauli susceptibility reported in the metallic regime of SbF$_{5}$-doped polyparaphenylene \cite{9} and the absence of any electron spin resonance signal in electrochemically cycled conducting polypyrrole \cite{6}. 

A bipolaron is spinless with an electric charge $\pm$2$e$. It is then thought of as analogous to the Cooper pair in the BCS theory of superconductivity \cite{10}, which consists of two electrons coupled through a lattice vibration, $i.e.$, a phonon. The formation of a bipolaron implies that the energy gained by the interaction with the lattice is larger than the Coulomb repulsion between the two charges of same sign confined in the same site \cite{11}. There has been speculation that organic polymeric conductors could become superconductors. The similarities between bipolaron and Cooper pair in the BCS theory of superconductivity make this prospect even more interesting. The natural separation of spin and charge in such one-dimensional systems remains an active area of condensed matter research and is thought to be relevant to high temperature superconductivity. Furthermore, the charge pairing (bipolaron) can be formed at room temperature in conducting polymers due to strong electron$-$lattice interaction \cite{7,8}.  The strong electron-lattice interaction can also provide a high energy scale for these systems. All these features make the polymer conductors as the promising candidates for high$-$temperature superconductors. 

Among conducting polymers, polyparaphenylene is attractive for its non-degenerate ground state structure, high conductivity \cite{12}, bipolaron conducting mechanism \cite{4,5,6,13}, desirable stability in air even at elevated temperatures, the absence of isomerization processes due to doping, as well as the formation of rechargeable batteries \cite{14}. In view of these facts, exploring possible superconductivity in conducting polyparaphenylene is the goal of this study. $p$-Oligophenyls possess the similar molecular structure of polyparaphenylene but with a shorter chain. For simplicity, the starting member of $p$-oligophenyls $-$ $p$-terphenyl with three phenyl rings is chosen for such a purpose. By doping potassium into $p$-terphenyl, we observe a superconducting transition at temperature of 7.2 K in this compound. The realization of superconductivity in $p$-terphenyl not only opens a window for finding superconductors in conducting polymers but also highlights the important role of bipolarons probably played for superconductivity. 

Potassium-doped $p$-terphenyl was synthesized from purified $p$-terphenyl (99.5\%) and potassium metal (99\%) purchased from Sigma-Aldrich and Sinopharm Chemical Reagent, respectively, according to the procedure reported previously \cite{pre15}. The nature of the carriers of this material was identified by Raman scattering measurements, which were performed on an in-house system with Charge Coupled Device and Spectrometer from Princeton Instruments in a wavelength of 660 nm and power less than 1 mW to avoid possible damage of samples. $p$-Terphenyl consists of benzene rings linked in the para position. Crystallographic data of this material indicate that the carbon-carbon bond lengths within the rings are about 1.40 \AA\/ and those between rings are about 1.51 \AA\/ \cite{15}. In the solid state, two successive benzene rings are tilted with respect to each other by about 23$^{o}$, which is induced by the steric repulsion between hydrogen atoms in ortho positions. Upon doping, two potassium atoms intercalate in the two carbon-carbon bonds between rings. The chain becomes nearly coplanar with almost flat titling angle for benzene rings. Along the chain, the C$-$C bond between rings is remarkably reduced. Parallel bonds in the rings are decreased to acquire a more pronounced double-bond character, while inclined bonds increase significantly. As a result, the inner ring has a strong quinoidal character \cite{4}. These geometric modifications bring about dramatic difference of the Raman spectra compared to the characteristic bands of the pristine material (Fig. 1). Five regions of Raman active modes from the low to high frequencies correspond to the lattice, C$-$C$-$C bending, C$-$H bending, C$-$C stretching, and C$-$H stretching modes. All these modes were observed in pristine $p$-terphenyl, which is consistent with previous works \cite{16,17}. The obtained Raman spectra and features for potassium-doped $p$-terphenyl are in good agreement with those in samples synthesized by various methods \cite{17}. 

Upon doping potassium into $p$-terphenyl, the strongest intra-ring C$-$C stretching bands at 1605 and 1593 cm$^{-1}$ show little downward shifts in wavenumber. They merge to bipolaronic bands centered at 1589 cm$^{-1}$ \cite{18,19}. The downshifts are the result of the increase of inclined C$-$C bond lengths within the rings due to the doping. By contrast, the 1275 cm$^{-1}$ band in the pristine is corrected with the 1291 and 1348 cm$^{-1}$ bands in the doped sample \cite{20}. These bands are mainly attributed to the inter-ring C$-$C stretching vibration. The observed upshifts with doping mainly reflect the length decrease of the C$-$C bonds between rings. Therefore, the increase of $\pi$-€"bond order of the inter-ring C$-$C bond is expected. These effects due to the formation of bipolarons drive the structural change of the molecule from the benzenoid to quinoid. The two bipolaronic bands at 1168 and 1219 cm$^{-1}$ correspond to the 1222 cm$^{-1}$ band in the pristine. The former (latter) can be assigned the C$-$H bending of external (internal) rings \cite{19}. The 1222 cm$^{-1}$ band arises from a mode concentrated in the inner region of the molecule with no contribution from the terminal rings \cite{16}. The rigidity of this band indicates that the rings of the molecule maintain well after the formation of bipolarons. The strong band at 1473 cm$^{-1}$ and the triple bands centered at 985 cm$^{-1}$ are all from new bipolarons. The Raman active 1473 cm$^{-1}$ mode is due to the vibration of the C$-$H bending of external rings \cite{19}. The appearance of this mode can be considered as the fingerprint for the newly formed bipolarons. The triple bands are from the in-plane ring bending \cite{19}. These new bands result from the loss of translational symmetry due to the formation of the bipolarons in the polymer chain. Our Raman spectra provide rather clear and solid evidence for the formation of bipolarons in our sample.

\begin{figure}[tbp]
\centerline{\includegraphics[width=\columnwidth]{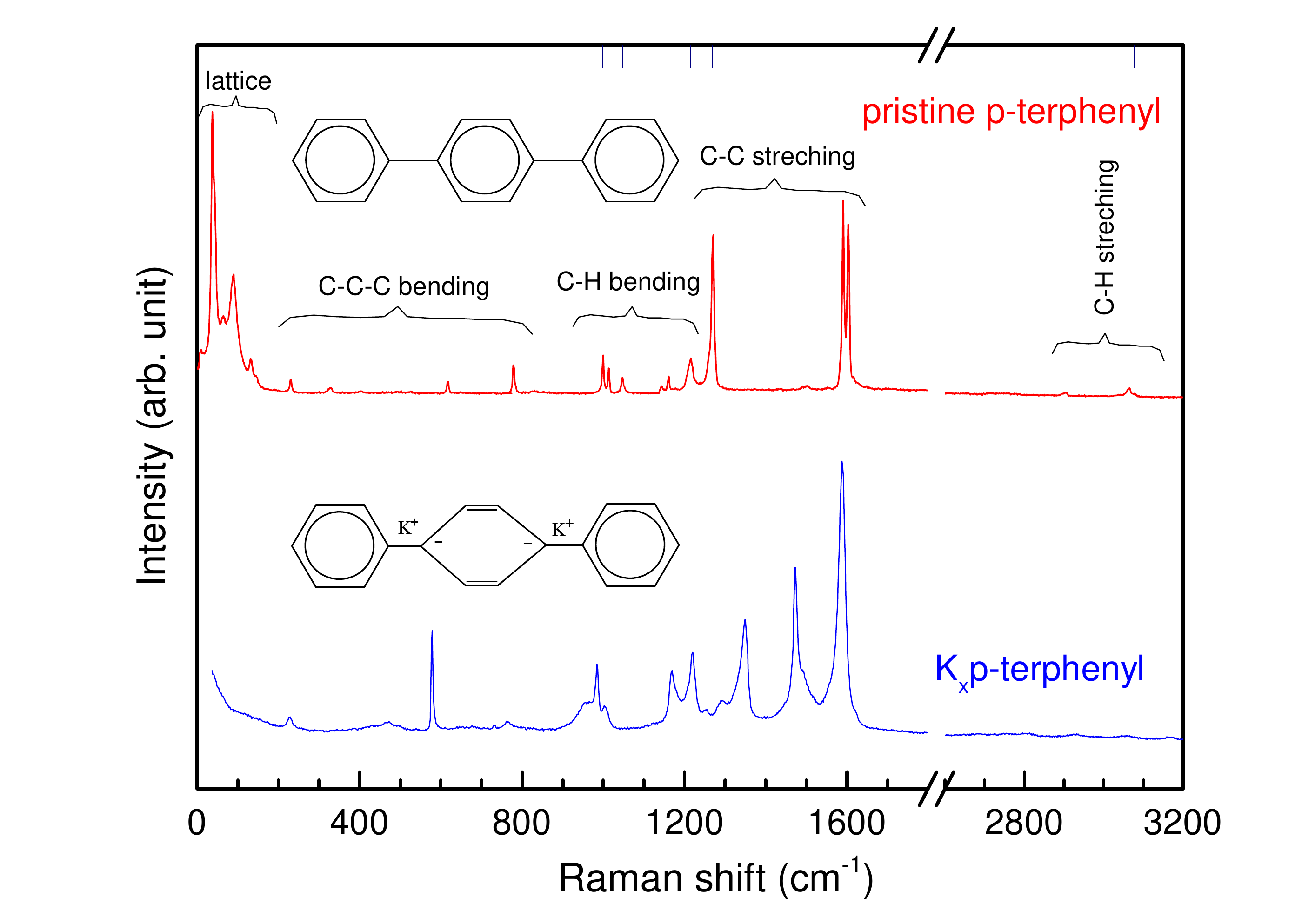}}
\caption{(Color online) Raman scattering spectra of pristine $p$-terphenyl and potassium-doped $p$-terphenyl collected at room temperature.  Upper left presents the molecular structure of $p$-terphenyl. Lower left shows its modification upon dopant. The sticks in the upper horizontal axis give the peak positions of the vibrational modes in pristine material. }
\end{figure}

Having clarified the bipolaronic character, we turn to examine the possible superconductivity in the synthesized sample. Zero electrical dc resistivity and Meissner effect are the two essential characters for a material to be called as a superconductor. Both of them are reached at a critical or superconducting transition temperature $T_{c}$ and below. However, the temperature$-$dc conductivity measurements in these materials are not as easy as one thought. They are mechanically brittle, so it is very difficult to fabricate reliable electrical contacts to measure their intrinsic transport properties. In fact, their true metallic state has not been achieved from the temperature$-$dc conductivity measurements \cite{13,21} but was determined mainly by optical \cite{13,21} and magnetic \cite{22,23} techniques. However, this difficulty can be safely removed in detecting superconductivity by using magnetic susceptibility technique. As a bulk probe technique, both the dc and ac magnetic susceptibility measurements enable the Meissner effect to be well characterized. In the ac magnetic susceptibility measurement, the real component $\chi^{\prime}$ of the susceptibility is a measure of the magnetic shielding and the imaginary component $\chi^{\prime\prime}$ is a measure of the magnetic irreversibility. The nearly zero $\chi^{\prime\prime}$ in the superconducting state can be considered as the signature of zero resistivity. Such zero-resistivity feature of $\chi^{\prime\prime}$ has been observed for various unconventional superconductors \cite{24}. In fact, many other superconducting properties such as the resistivity, critical temperature and fields, London and Campbell penetration depths, critical current density, granularity and content of superconducting phase, irreversibility line, and pinning potential can be also obtained from the ac data \cite{25}. 

\begin{figure}[tbp]
\centerline{\includegraphics[width=\columnwidth]{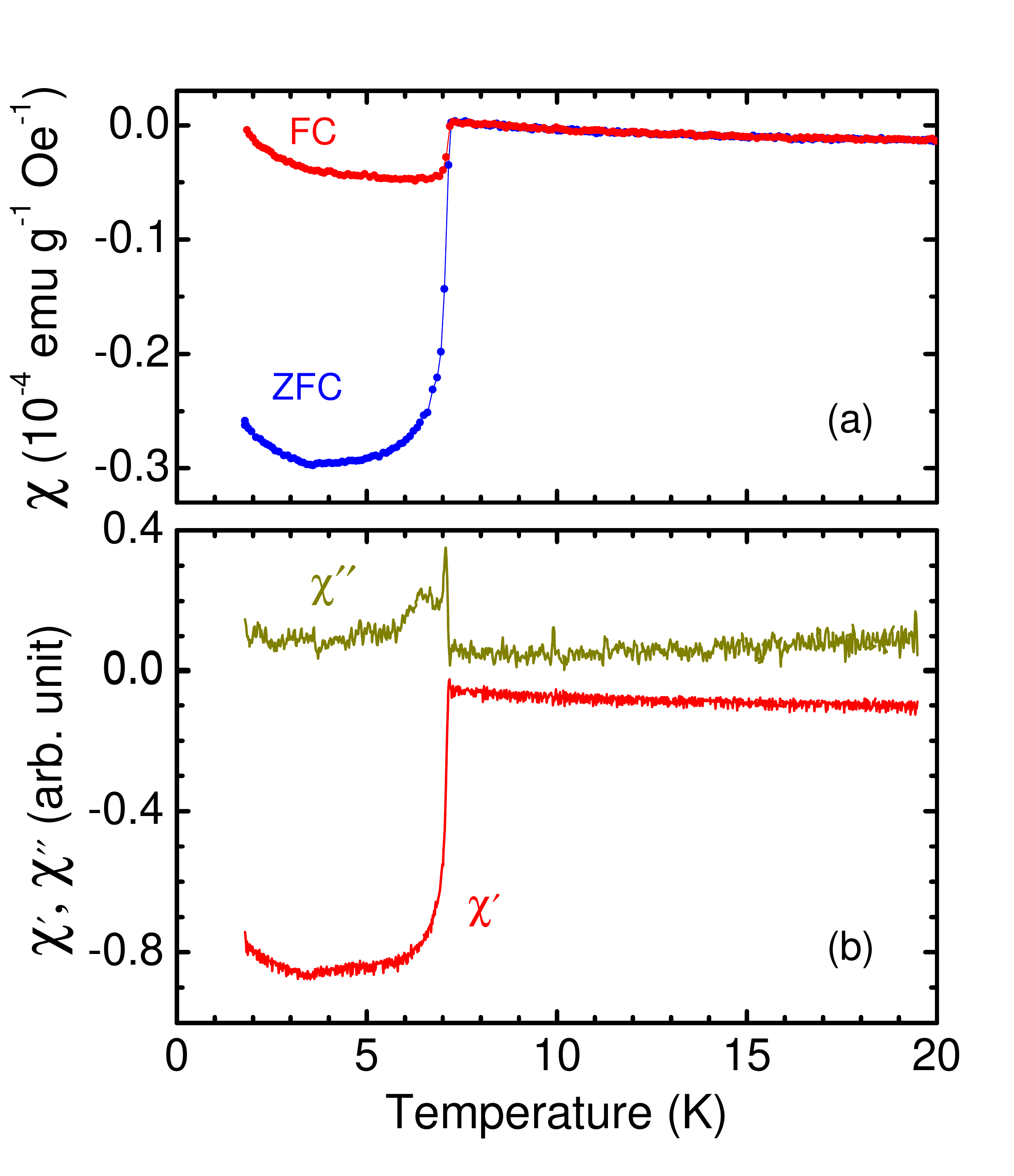}}
\caption{(Color online) (a) The temperature dependence of the dc magnetic susceptibility $\chi$ for potassium-doped $p$-terphenyl in the applied magnetic field of 10 Oe with field cooling (FC) and zero-field cooling (ZFC).  (b) Real ($\chi^{\prime}$) and imaginary ($\chi^{\prime\prime}$) components of the ac magnetic susceptibility as a function of temperature. The probe harmonic magnetic field and frequency are 5 Oe and 234 Hz, respectively. }
\end{figure}

The measurements of both the dc and ac magnetic susceptibilities were performed on our sample with a SQUID magnetometer (Quantum Design MPMS3) in the temperature range of 1.8-300 K. Figure 2(a) shows the dc magnetic susceptibility $\chi$ in the applied magnetic field of 10 Oe with field cooling (FC) and zero-field cooling (ZFC) in the temperature range of 1.8$-$20 K. One can clearly see that $\chi$ shows a sudden decrease below 7.2 K in the ZFC run, while it exhibits a small upturn after the drop in the FC run. Both decreases yield a superconducting transition temperature $T_c$=7.2 K. This shape of the magnetization susceptibility curve is consistent with the well-defined Meissner effect. This feature together with narrow transition width indicates good superconducting properties of this sample. 

\begin{figure}[tbp]
\centerline{\includegraphics[width=\columnwidth]{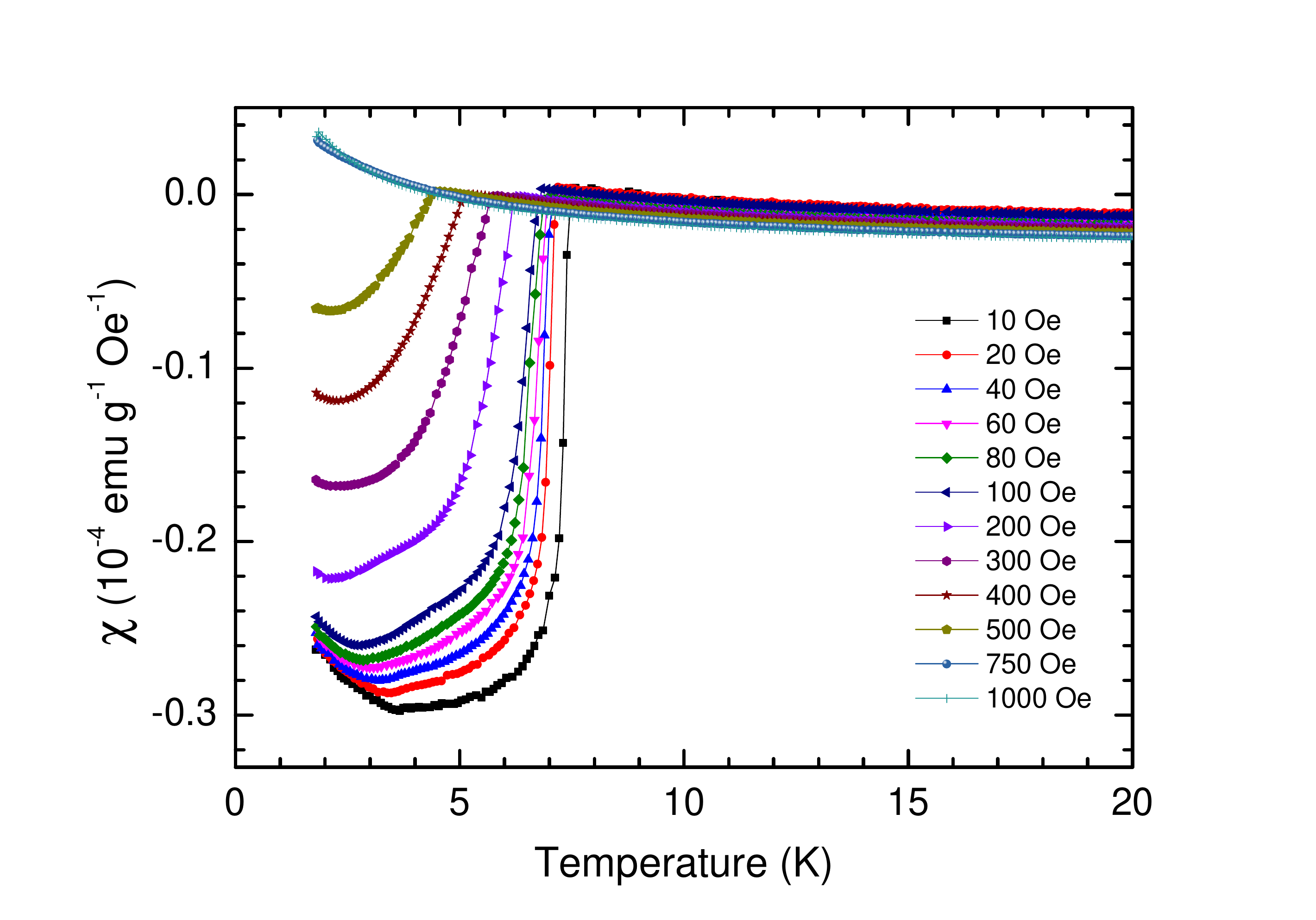}}
\caption{(Color online) The temperature dependence of the dc magnetic susceptibility of potassium-doped $p$-terphenyl measured at various magnetic fields up to 1000 Oe in the zero-field-cooling (ZFC) run. }
\end{figure}

Figure 2(b) presents the temperature dependence of the real $\chi^{\prime}$ and imaginary $\chi^{\prime\prime}$ components of the ZFC ac susceptibility. Note that a bulk superconducting state is again observed from this ac technique. Two inflection anomalies occur upon cooling at the exactly same temperature of 7.2 K, which coincides with the $T_{c}$ value already determined from the dc technique. As can be seen, both $\chi^{\prime}$ and $\chi^{\prime\prime}$ are nearly the constants close to zero above the transition because there is no flux exclusion in the normal state. In the superconducting state below 7.2 K, the diamagnetic behavior leads to a negative $\chi^{\prime}$ which becomes more negative as temperature is reduced and more flux is expelled from the sample. Here the flux penetrating the sample lags the external flux, resulting in the dissipation seen in the $\chi^{\prime\prime}$ signal. The peak in $\chi^{\prime\prime}$ occurs when the flux is just penetrating as far as the center of the sample. The flatness of $\chi^{\prime}$ in the maxed state rules out the possibility for the existence of superconducting grains or the second superconducting phase in our sample \cite{25}. The nearly zero $\chi^{\prime\prime}$ after passing a dome upon cooling implies the reach of zero-resistivity in the superconducting state.  

\begin{figure}[tbp]
\centerline{\includegraphics[width=\columnwidth]{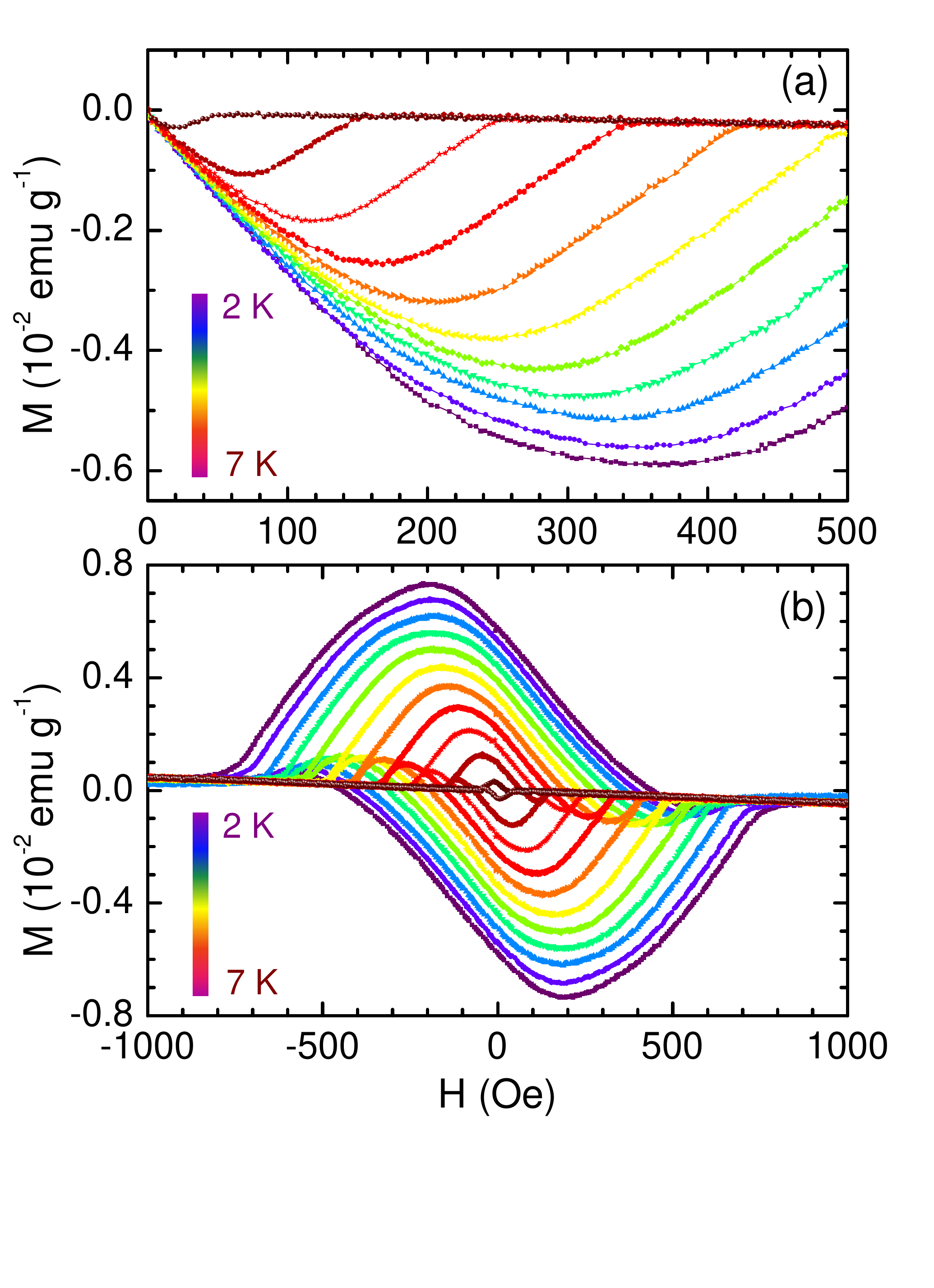}}
\caption{(Color online) The magnetic field dependence of the magnetization of potassium-doped $p$-terphenyl at various temperatures in the superconducting state measured in the zero-field-cooling (ZFC) run up to 500 Oe. The lower critical field $H_{c1}$ is marked by an arrow defined by the deviation from the linear $M$ vs $H$ behavior. (b) The magnetization loop with scanning magnetic field up to 1000 Oe measured at various temperatures in the superconducting state. }
\end{figure}

The obtained superconductivity in potassium-doped $p$-terphenyl is further supported by the evolution of the $\chi-T$ curve with the applied magnetic fields (Fig. 3). The curve gradually shifts towards the lower temperature with increasing magnetic field. The superconducting fraction is thus suppressed by the applied magnetic field. These features demonstrate that the obtained superconductivity is intrinsic for the studied material. For the magnetic fields larger than 750 Oe, the $\chi-T$ curve moves upwards and the superconducting transition is hardly observed within the lowest temperature measured down to 2 K.

The Meissner effect of this material is clearly demonstrated from the magnetization measurements as a function of magnetic field up to 500 Oe in the ZFC run at various temperatures in the superconducting state. The results are summarized in Fig. 4(a). The linearity of the $M-H$ curves at these selected temperatures is characteristic for the Meissner phase of this superconductor, yielding different $H_{c1}^{\prime}$s for corresponding temperatures based on the high field deviation from the linear behavior. An increase of the linearity region with decreasing temperature indicates that the lower temperature corresponds to the higher $H_{c1}$. The Meissner effect becomes strong when temperature is far away from $T_{c}$ of 7.2 K.

Figure 4(b) shows the magnetization loop with magnetic field up to 1000 Oe measured at various temperatures between 2 and 7 K in the superconducting state. The hysteresis loop along the two opposite magnetic field directions provides evidence for the type-II superconductor. Its clear diamond-like shape is a typical character for a superconductor. The asymmetry along the vertical $H=$0 axis provides evidence for the wide-band susceptibilities \cite{25}. The diamond expands from the inner to the outer with lowering the temperature, yielding a higher upper critical field $H_{c2}$ in the horizontal axis for a lower temperature. 

\begin{figure}[tbp]
\centerline{\includegraphics[width=\columnwidth]{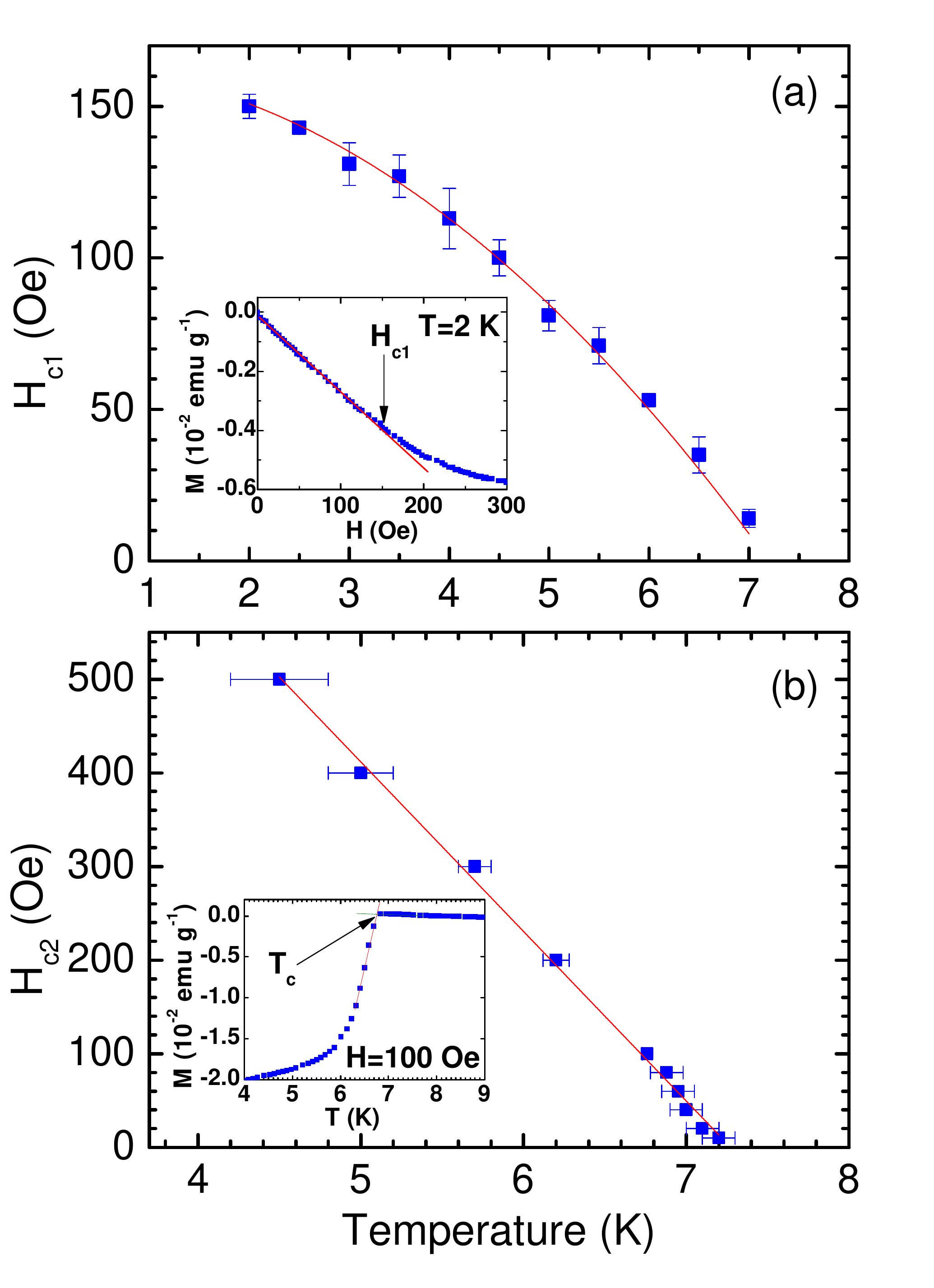}}
\caption{(Color online) (a) The temperature dependence of the lower critical field $H_{c1}$($T$). Error bars represent estimated uncertainty in determining $H_{c1}$. The solid line represents the empirical law $H_{c1}$($T$)/$H_{c1}$(0) $=$1$-$($T/T_{c}$)$^{2}$. Inset: The field-dependent magnetization at temperature of 2 K gives the determination of $H_{c1}$. (b) The temperature dependence of the upper critical field $H_{c2}$($T$). The error bars represent the uncertainty in the rounding of the transition. The line is the fitting to the Werthamer-Helfand-Hohenberg theory. Inset: The temperature-dependent magnetization at the applied field of 100 Oe displays the definition of $T_{c}$. }
\end{figure}

Figure 5 shows the temperature dependences of the critical fields of this 7.2 K superconductor. The inset of Fig. 5(a) shows how $H_{c1}$ is determined for a given temperature based on the deviation of the linear behavior at higher field.  The zero-temperature extrapolated value of $H_{c1}$(0) is 163$\pm$2 Oe. Hc2 can be determined from the $M-T$ curves measured at various magnetic fields. Here, $T_{c}$ for a given field is determined from the intercept of linear extrapolations from below and above the transition [Inset of Fig. 5(b)]. The obtained $H_{c2}$($T$) as a function of temperature is shown in Fig. 5(b). Using the Werthamer-Helfand-Hohenberg formula \cite{26}, the zero-temperature $H_{c2}$(0) =1317$\pm$22 Oe is obtained. From $H_{c2}$(0) and $H_{c1}$(0), we evaluate the zero-temperature superconducting London penetration depth $\lambda_{L}$ and Ginzburg-Landau coherence length $\xi_{GL}$ using the expressions $H_{c2}$(0)= $\Phi_{0}/2\pi\xi_{GL}^{2}$ and $H_{c1}$(0)$=(\Phi_{0}/4\pi\lambda_{L}^{2})\ln(\lambda_{L}/\xi_{GL})$ \cite{27} with the flux quantum $\Phi_{0}=2.0678\times10^{-15}$ Wb.  Substituting the obtained $H_{c2}$(0) and $H_{c1}$(0), we obtain $\xi_{GL}=500\pm4$ \AA\/ and $\lambda_{L}=760\pm3$ \AA\/ for this new superconductor. Thus the Ginzburg-Landau parameter $\kappa=\lambda_{L}/\xi_{GL}$=1.52 is obtained for this superconductor, a value close to some cuprate superconductors. 

The present results provide unambiguous evidence for superconductivity in $p$-terphenyl, which was driven by the exploration of bipolaronic superconductivity in conducting polymers.  The formation of bipolarons in this material has been identified from Raman scattering measurements. Although the mechanism of superconductivity either from bipolarons \cite{28,29} or Cooper pairing \cite{10} has not been pinned down yet and remains a subject for future research, the electron-lattice (phonon) coupling should be the key to the understanding of the observed superconductivity. Since the coupling strength can be finely tuned in such systems, the $T_{c}$ enhancement or finding materials with higher $T_{c}^{\prime}$s is expected. Nowadays conducting polymers can be manufactured in industry. Many advantages of these cheap materials make the technology applications unbelievably wide if they become superconductors. In the light of the wealth of unanticipated discoveries that the field of polymeric conductors has witnessed in the past nearly 40 years, the scientific and technological future of conducting polymers as superconductors has to be viewed optimistically.

\begin{acknowledgments}
This work was supported by the Natural Science Foundation of China. We thank Hai-Qing Lin and Ho-Kwang Mao for strong support and valuable discussion. 
\end{acknowledgments}


\begin{thebibliography}{99}

\bibitem{1} {\it Handbook of Conducting Polymers}, edited by T. A. Skotheim (Dekker, New York, 1986).

\bibitem{2} A. J. Heeger, S. Kivelson, J. R. Schrieffer, and W. P. Su, Rev. Mod. Phys. \textbf{60}, 781 (1988).

\bibitem{3} J. L. Br$\acute{e}$das, R. R. Chance, and R. Silbey, Phys. Rev. B \textbf{26}, 5843 (1982).

\bibitem{4} J. L. Br$\acute{e}$das, B. Th$\acute{e}$mans, J. G. Fripiat, J. M. Andr$\acute{e}$, and R. R. Chance, Phys. Rev. B \textbf{29}, 6761 (1984).

\bibitem{5} R. R. Chance, J. L. Br$\acute{e}$das, and R. Silbey, Phys. Rev. B \textbf{29}, 4491 (1984).

\bibitem{6} J. C. Scott, P. Pfluger, M. T. Krounbi, and G. B. Street, Phys. Rev. B \textbf{28}, 2140 (1983).

\bibitem{7} J. H. Kaufman, N. Colaneri, J. C. Scott, and G. B. Street, Phys. Rev. Lett. \textbf{53}, 1005 (1984).

\bibitem{8} G. Crecelius, M. Stamm, J. Fink, and J. J. Ritsko, Phys. Rev. Lett. \textbf{50}, 1498 (1983).

\bibitem{9} M. Peo, S. Roth, K. Dransfeld, B. Tieke, J. Hocker, H. Gross, A. Grupp, and H. Sixl, Sold State Commun. \textbf{35}, 119 (1980).

\bibitem{10} J. Bardeen, L. N. Cooper, and J. R. Schrieffer, Phys. Rev. \textbf{108}, 1175 (1957).

\bibitem{11} P. W. Anderson, Phys. Rev. Lett. \textbf{34}, 953 (1975).

\bibitem{12} L. W. Shacklette, R. R. Chance, D. M. Ivory, G. G. Miller, and R. H. Baughfnan, Synth. Met. \textbf{1}, 307 (1979). 

\bibitem{13}P. Kuivalainen, H. Stubb, H. Isotalo, P. Yli-Lahti, and C. Holmstr$\ddot{o}$m, Phys. Rev. B \textbf{31}, 7900 (1985).
 
\bibitem{14} L. W. Shacklette, R. L. Elsenbaumer, R. R. Chance, J. M. Sowa, D. M. Ivory, G. G. Miller, and R. H. Baughrnan, J. Chem. Soc. Chem. Commun. \textbf{1982}, 361 (1982).

\bibitem{pre15} P. Balk, G. J. Hoijtink, and J. W. H. Schreurs, Rec. Trav. Chim. \textbf{76}, 813 (1957).

\bibitem{15} J. L. Baudour, H. Cailleau, and W. B. Yelon, Acta Crystallogr. Sect. B \textbf{33}, 1773 (1977).

\bibitem{16} H. Ohtsuka, Y. Furukawa, and M. Tasumi, Spectrochim. Acta A \textbf{49}, 731 (1993).

\bibitem{17} L. O. P$\acute{e}$res, M. Spiesser, and G. Froyer, Synth. Met. \textbf{155}, 450 (2005). 

\bibitem{18} A. Simonneau, G. Froyer, J.-P. Buisson, and S. Lefrant, Synth. Met. \textbf{84}, 627 (1997). 

\bibitem{19} M. Dubois, G. Froyer, G. Louarn, and D. Billaud, Spectrochim. Acta A \textbf{59}, 1849 (2003).

\bibitem{20} Y. Furukawa, H. Ohtsuka, and M. Tasumi, J. Raman Spectroscopy \textbf{24}, 551 (1993).

\bibitem{21} C. K. Chiang, C. R. Fincher, Jr., Y. W. Park, A. J. Heeger, H. Shirakawa, E. J. Louis, S. C. Gau, and A. G. MacDiarmid, Phys. Rev. Lett. \textbf{39}, 1098 (1977).

\bibitem{22} M. Sebti, Merlin, J. Ghanbaja, and D. Billaud, Synth. Met. \textbf{84}, 665 (1997).

\bibitem{23} A. C. Kolbert, S. Caldarelli, K. F. Thier, N. S. Sariciftci, Y. Cao, and A. J. Heeger, Phys. Rev. B \textbf{51}, 1541 (1995).

\bibitem{24} E. B. Yagubskii, N. D. Kushch, A.V. Kazakova, L. I. Buravov, V. N. Zverev, A. I. Manakov, S. S. Khasanov, and R. P. Shibaeva, JETP Lett. \textbf{82}, 99 (2005).

\bibitem{25} F. G$\ddot{o}$m$\ddot{o}$ry, Supercond. Sci. Technol. \textbf{10}, 523 (1997).

\bibitem{26} N. R. Werthamer, E. Helfand, and P. C. Hohenberg, Phys. Rev. \textbf{147}, 295 (1966). 

\bibitem{27} M. Tinkham, {\it Introduction to Superconductivity} (McGraw-Hill, New York, 1975). 

\bibitem{28} B. K. Chakraverty, J. Physique \textbf{42}, 1351 (1981).

\bibitem{29} A. Alexandrov and J. Ranninger, Phys. Rev. B \textbf{24}, 1164 (1981).


\end{thebibliography}
\end{document}